\documentclass[aps,prl,%twocolumn,
showpacs,superscriptaddress,groupedaddress]{revtex4}  % for review and submission
\usepackage{graphicx}  % needed for figures
\usepackage{dcolumn}   % needed for some tables
\usepackage{bm}        % for math
\usepackage{amssymb}   % for math

\begin{document}

% The following information is for internal review, please remove them for submission
%\widetext
%\leftline{Version xx as of \today}
%\leftline{Primary authors: Joe E. Physics}
%\leftline{To be submitted to (PRL, PRD-RC, PRD, PLB; choose one.)}
%\leftline{Comment to {\tt d0-run2eb-nnn@fnal.gov} by xxx, yyy}
%\centerline{\em D\O\ INTERNAL DOCUMENT -- NOT FOR PUBLIC DISTRIBUTION}

% the following line is for submission, including submission to the arXiv!!
%\hspace{5.2in} \mbox{Fermilab-Pub-04/xxx-E}

\title{ Generate tensor network state by sequential single-photon scattering in waveguide QED systems}
%\input author_list.tex       % D0 authors (remove the first 3 lines
                             % of this file prior to submission, they
                             % contain a time stamp for the authorlist)
                             % (includes institutions and visitors)
\author{Shanshan Xu}
\email{xuss@stanford.edu}
\affiliation{Department of Physics, Stanford University, Stanford, California 94305}

\author{Shanhui Fan}
\email{shanhui@stanford.edu} \affiliation{Department of Electrical Engineering, Ginzton Laboratory, Stanford University, Stanford, California 94305}

\date{\today}

\begin{abstract}
We propose a scheme to generate photonic tensor network states by sequential scattering of photons in waveguide QED systems. We show that sequential
scatterings can convert a series of unentangled photons into any type of matrix product states. 
We also demonstrate the possibility of generating
projected entangled pair states  with arbitrary network representation by photon re-scattering.

\end{abstract}

%\pacs{}
\maketitle

%\section{\label{sec:level1}First-level heading}
% sections are not used for PRL papers

Quantum tensor network states with a local entanglement structure, including matrix product states (MPS) \cite{pgvwc}  and its generalization 
projected entangled pair states (PEPS) \cite{vc,vwpgc}, have become a key tool in quantum many-body physics \cite{vmc,orus}. 
It has been observed that for a large class of local Hamiltonians, their ground states can be well approximated by these states \cite{hastings06, hastings07,fwbse}.
This observation provides the foundation of various classical algorithms designed to efficiently simulate
strongly correlated quantum systems \cite{dmrg,scholl,scholl2,vidal,vidal2}. 
In addition to their theoretical interests, MPS and PEPS
have a variety of practical applications in quantum information science including quantum computing \cite{rb}, quantum key distribution \cite{grtz} or quantum cryptography \cite{ekert, bbgm}.
Therefore, there have been significant interests in generating such states.
Since photons are ideal carriers of quantum information for their hight transfer rates and long coherence times, 
there have been significant works for generating MPS or PEPS for photons using cavity quantum electrodynamics (QED) systems by having such systems emitting photons sequentially \cite{ssvcw,shwcs,lr,scsdgkld,pczl}. 
However, the types of states that can be generated in such cavity QED schemes are limited to the one-dimensional MPS as well as a particular type of PEPS with 
square lattice networks \cite{pczl}.

\begin{figure}[h]
\includegraphics
[width=0.8\textwidth] {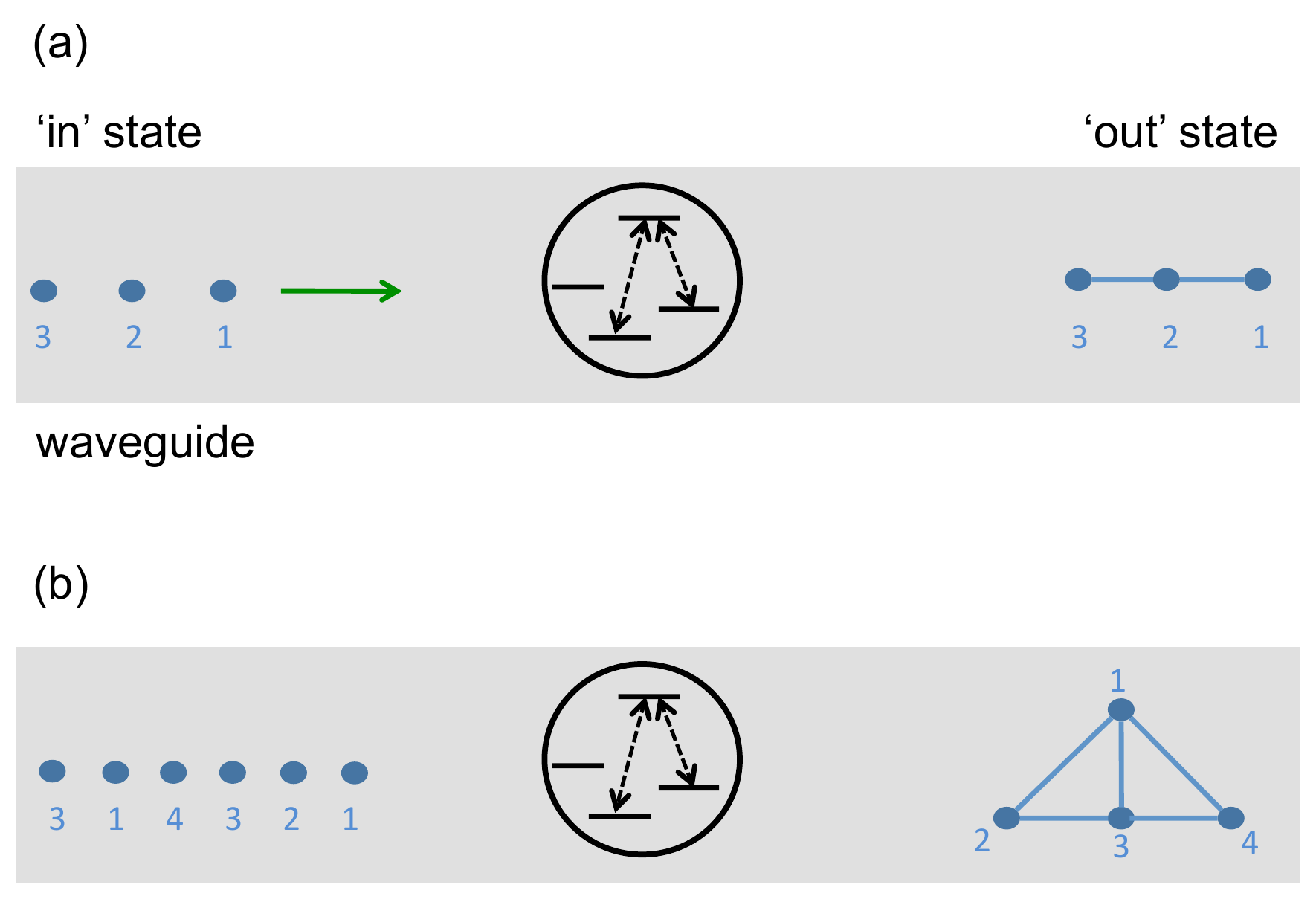} \caption{ Generation of MPS and PEPS by
 sequential scatterings of photons in a waveguide coupled to a multi-level atom. (a) Three untangled photons sequentially scatter against the atom. The output state is a three-photon MPS.
 (b) Four unentangled photons sequentially sent in and scatter against the atom. After the fourth photon scattering, we do second scatterings for the first and third photons. The output state
 is then a simple two-dimensional PEPS.
} \label{fig1}
\end{figure}

In this letter, we propose an alternative approach to generate photonic tensor network states by sequentially scattering photons against a local
quantum multi-level system in a waveguide QED system.  
For succinctness, for the rest of the paper we refer to such a local quantum multi-level system as an `atom'. 
In practice, such an `atom' can be a real atom, or a solid state multi-level system such as a quantum dot, a color center or a superconducting qubit. 
We prove that we can generate arbitrary one-dimensional MPS states with their bond dimension equal to the dimension of the Hilbert space of the atom
(Fig.~\ref{fig1}(a)). 
Moreover, unlike the cavity QED approach, in this approach there is an additional flexibility of photon re-scattering, i.e. the scattering of the same photons multiple times against the atom. 
We show that such a flexibility can enable one to generate PEPS with arbitrary graph representation (Fig.~\ref{fig1}(b)). 

As a start, we consider a waveguide QED system where photons propagate along the waveguide and scatter against an atom supporting multiple ground states. 
The scattering process in general can be described by the single-photon scattering matrix (S matrix) with its matrix element denoted by
\begin{eqnarray}\label{sS}
S_{\beta\alpha}^{ji} = \langle j,\beta | \,\hat{S} \,|  i, \alpha\rangle \,.
\end{eqnarray}
In this letter we use Greek letters such as $\alpha, \beta$ to denote the atom's ground states and Latin letters such as $i,j$ to denote the internal degrees of freedom of the photon,
which can be either polarizations or discrete frequency bins of photons. 
Such a waveguide QED system has been extensively discussed in the literature \cite{sfoptic,crdl,sf,sfA,ss,ll,zgb2,fks,sfs,lsb,eks,gmmtg,dr,koz,zgb,rf,jg,zb,xf,ll2,sf2,r,rf2,szgm,lhl,xfa,cmsdcc,scc}. We show a concrete example of one such waveguide QED system
including its Hamiltonian and single-photon S matrix in the later part.  
For most of the paper we proceed generally by only assuming that such an S matrix exists. 
For a system described by the single-photon S matrix (\ref{sS}), if we scatter a photon in an arbitrary input state $\sum_{i} d^i|i\rangle$
against the atom in the state $\sum_{\alpha} I_{\alpha}|\alpha\rangle$, the output state in general is an entangled state between the photon and the atom having the form:
\begin{eqnarray}
\hat{S}\, \left[\sum_{i} d^i|i\rangle \otimes\sum_{\alpha} I_{\alpha}|\alpha\rangle\right] = \sum_{i,\alpha}  d^i  I_{\alpha}\hat{S}|i,\alpha\rangle
= \sum_{i,\alpha}   d^i  I_{\alpha}\sum_{j,\beta } S_{\beta\alpha}^{ji}  |j,\beta\rangle\,,\nonumber
\end{eqnarray}
or in short 
\begin{eqnarray}\label{sb}
\hat{S}\, \left[  d^i|i\rangle\otimes I_{\alpha}|\alpha\rangle\right] =   S_{\beta\alpha}^{ji} d^i  I_{\alpha} |j,\beta\rangle\,.
\end{eqnarray}
In (\ref{sb}) and also for the rest of the paper, we use the Einstein summation convention to simplify the expression.

For the system as described by (\ref{sS}) and (\ref{sb}), we first present a protocol to generate MPS by sequentially scattering photons against the 
atom. Suppose the atom is initially prepared to be in the state $I_{\alpha_0}|\alpha_0\rangle$ and against this atom we sequentially scatter $n$ photons in the states $d^{[1]j_1}|j_1\rangle, \cdots, 
d^{[n]j_n}|j_n\rangle$, respectively. As shown in (\ref{sb}), after the first scattering, the system is in the state $S_{\beta_1\alpha_0}^{i_1j_1} d^{[1]j_1}  I_{\alpha_0} |i_1,\beta_1\rangle$.
We then perform a unitary operation $\hat{R}^{[1]}\equiv R^{[1]}_{\alpha_1\beta_1}|\alpha_1\rangle\langle \beta_1|$ on the atom to arrive at the state
$R^{[1]}_{\alpha_1\beta_1}S_{\beta_1\alpha_0}^{i_1j_1} d^{[1]j_1}  I_{\alpha_0} |i_1,\alpha_1\rangle$.
Repeating this process by scattering a second photon against the atom followed by a unitary operation $\hat{R}^{[2]}$ on the atom, we obtain the state 
$R^{[2]}_{\alpha_2\beta_2}S_{\beta_2\alpha_1}^{i_2j_2} d^{[2]j_2}\,R^{[1]}_{\alpha_1\beta_1}S_{\beta_1\alpha_0}^{i_1j_1} d^{[1]j_1}  I_{\alpha_0} |i_2i_1, \alpha_2\rangle$. 
This process can be repeated for all the other photons. 
 Finally, as the very last step, we disentangle the photons from the atom by projecting the atomic state to some final state $F_{\alpha_n}|\alpha_n\rangle$.
The resulting $n$-photon state is of the form 
\begin{eqnarray}\label{mps1}
|\psi_n\rangle =F^*_{\alpha_n} R^{[n]}_{\alpha_n\beta_n}S_{\beta_n\alpha_{n-1}}^{i_nj_n} d^{[n]j_n}\,\cdots\,
R^{[2]}_{\alpha_2\beta_2}S_{\beta_2\alpha_1}^{i_2j_2} d^{[2]j_2}\,R^{[1]}_{\alpha_1\beta_1}S_{\beta_1\alpha_0}^{i_1j_1} d^{[1]j_1}  I_{\alpha_0} |i_n\cdots i_2 i_1\rangle\,.
\end{eqnarray} 
We define a series of matrices $\mathbf{A}^{[k]i_k}$ for $1\leq k \leq n$ with entries 
\begin{eqnarray}\label{Amat}
{A}^{[k]i_k}_{\alpha_k \alpha_{k-1}} \equiv R^{[k]}_{\alpha_k\beta_k}S_{\beta_k\alpha_{k-1}}^{i_kj_k} d^{[k]j_k}\,,
\end{eqnarray}
and represent the initial and final states of the atom as $\mathbf{I}$ and $\mathbf{F}$ with entries $I_{\alpha_0}$ and $F_{\alpha_n}$, respectively, 
 the state (\ref{mps1}) can be written more transparently as 
 \begin{eqnarray}
 |\psi_n\rangle = F^*_{\alpha_n} {A}^{[n]i_n}_{\alpha_n \alpha_{n-1}}\,\cdots\, {A}^{[2]i_2}_{\alpha_2 \alpha_{1}}{A}^{[1]i_1}_{\alpha_1 \alpha_{0}} I_{\alpha_0} |i_n\cdots i_2 i_1\rangle
 =\mathbf{F}^{\dag} \mathbf{A}^{[n]i_n}\cdots \mathbf{A}^{[2]i_2}\mathbf{A}^{[1]i_1} \mathbf{I} |i_n\cdots i_2 i_1\rangle\,,\label{mps2}
 \end{eqnarray}
 which is exactly in the MPS form with the open boundary condition. The bond dimension of such MPS is equal to the number of the ground states of the atom. 
Furthermore, the unitarity constraint of S matrix $S^{li *}_{\gamma \alpha} S^{lj}_{\gamma\beta} = \delta^{ij}\delta_{\alpha\beta}$ gives the normalizations of the $A$ matrices:
$\mathbf{A}^{[n]i\dag}\mathbf{A}^{[n]i} = d^{j*}d^j$.

Eq.(\ref{Amat}) provides an explicit connection between MPS representation and the single-photon S matrix of a waveguide QED system. One one hand, given the 
S matrix and rotation operations, we know immediately the type of MPS that we can generate. To remove the non-uniqueness of MPS representation, one can always convert 
the MPS in Eq. (\ref{Amat}) to its canonical form \cite{canonical}.  On the other hand, for a given $\mathbf{A}$ matrix,
there are enough degrees of freedom to construct a S matrix that satisfies (\ref{Amat}). Therefore, we can generate any type of translation-invariant MPS by designing a waveguide QED system with the proper S matrix.  

\begin{figure}[h]
\includegraphics
[width=0.4\textwidth] {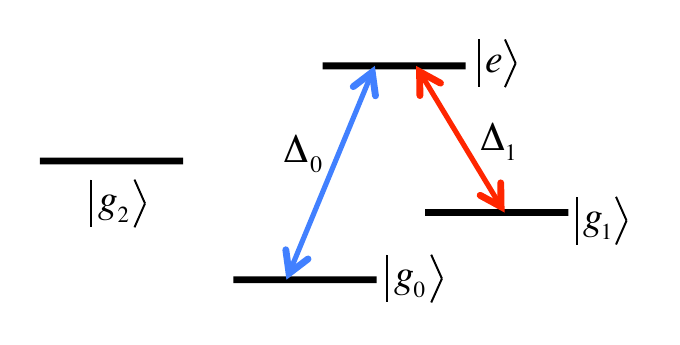} \caption{A four-level atom in the waveguide QED system used to generate either a W state or a GHZ state. The energy level $|g_2\rangle$ does not couple to the waveguide photons.
} \label{fig4}
\end{figure}

For illustration, here we provide a simple waveguide QED system and its S matrix that can be used to generate either a W-type state or a GHZ state by sequential scattering of photons. For three photons, these are the only two inequivalent ways of entanglement \cite{dvc}. Therefore, this construction provides an illustration that arbitrary MPS states can be indeed generated. 
The system consists of a four-level atom as shown in Fig.~\ref{fig4} and the Hamiltonian is described as
\begin{eqnarray}\label{int}
H&=&\int
dk\,k\,c_k^{\dag}\,c_k+ \sum_{\alpha=0}^2 \widetilde{\Delta}_{\alpha} |g_{\alpha}\rangle\langle g_{\alpha}|+\Omega |e\rangle\langle e|+\sum_{\alpha=0}^1\sqrt{\frac{\gamma_{\alpha}}{2\pi}}
\int dk \left(c_k^{\dag}\,|{\alpha}\rangle\langle e|+|e\rangle\langle {\alpha}|c_k\right)\,,
\end{eqnarray}
where $c_k\, (c_k^{\dag})$ is the annihilation (creation) operator
of the photon state in the waveguide. These operators satisfy the
standard commutation relation $[c_k, c_{k'}^{\dag}]=\delta(k-k')$.
 Here for simplicity we consider
a waveguide consisting of only a single mode in the sense of
Ref.\cite{sfA}. The argument here, however, can be straightforwardly
generalized to waveguides supporting multiple modes.
$\widetilde{\Delta}_0$, $\widetilde{\Delta}_1$ and $\Omega$ are the respective energy of the ground states $|g_0\rangle$, $|g_1\rangle$ and the excited state $|e\rangle$ of the atom satisfying
$\widetilde{\Delta}_0 <\widetilde{\Delta}_1<\Omega$.  We define ${\Delta}_{\alpha}\equiv \Omega-\widetilde{\Delta}_{\alpha}$ for $\alpha=0,1$.
The waveguide photons couple to both
$|g_0\rangle -|e\rangle$ and $|g_1\rangle -|e\rangle$ transitions of the atom with respective coupling constants $\sqrt{\gamma_0/2\pi}$ and $\sqrt{\gamma_1/2\pi}$. 
There is an additional energy level $|g_2\rangle$ that does not couple to any other atom levels by waveguide photons and thus does not appear
in the waveguide-atom coupling term in the Hamiltonian (\ref{int}).
This can be accomplished, for example, by choosing the frequency bins of the photons such that the photon frequency is off resonance from the transitions that involves $|g_2\rangle$.
The single-photon S matrix for this system is 
\begin{equation}\label{singleS}
\left[\mathbf{S}_{pk}\right]_{\beta\alpha}\equiv \langle p, g_{\beta}|S|k, g_{\alpha}\rangle = t_{\beta\alpha}(k)\, \delta(p-\Delta_{\beta}-k+\Delta_{\alpha})\,,\label{singleS}
\end{equation}
where $\alpha,\beta$ take values of $0,1$ and
\begin{equation}
t_{\beta\alpha}(k)=\delta_{\alpha\beta}-i\frac{\sqrt{\gamma_{\alpha}\gamma_{\beta}}}{k-\Delta_{\alpha}+i\left(\frac{\gamma_0}{2}+\frac{\gamma_1}{2}\right)}\label{tuv}
\end{equation}
is the transmission amplitude of the waveguide photon $|k\rangle$ when the initial and final states of the atom
are $|g_{\alpha}\rangle$ and  $|g_{\beta}\rangle$, respectively \cite{tl,ws,klyn,fb}. Furthermore, we only consider waveguide photons with two resonate frequencies $\Delta_0$ and 
$\Delta_1$. That is, we encode the photonic qubit in the frequency degree of freedom as $|i\rangle\equiv | \Delta_i\rangle$ for $i=0,1$.
Finally, let $\xi=\frac{\gamma_1-\gamma_0}{\gamma_0+\gamma_1}\,, \eta=-\frac{2\sqrt{\gamma_0\gamma_1}}{\gamma_0+\gamma_1}$
and $\left|\Delta_0-\Delta_1\right|\gg \gamma_0,\gamma_1$, the computed single-photon S matrix (\ref{singleS}) reduces to the form of (\ref{sS}) with
\begin{eqnarray}
\mathbf{S}^{00} = \left[\begin{array}{cc}\xi & 0 \\0 & 1\end{array}\right]\,,\,\,\,\,
\mathbf{S}^{01} = \left[\begin{array}{cc}0 & \eta \\0 & 0\end{array}\right]\,,\,\,\,\,
\mathbf{S}^{10} = \left[\begin{array}{cc}0 & 0 \\ \eta & 0\end{array}\right]\,,\,\,\,\,
\mathbf{S}^{11} = \left[\begin{array}{cc}1 & 0 \\0 & -\xi\end{array}\right]\,,
\end{eqnarray}
where the row and column index of the matrices are related to 
the atom's degrees of freedom $\alpha, \beta$. 
For an input state $|i,g_{\alpha}\rangle$, the output state remains the same when $i\neq \alpha$ but becomes  entangled when $i=\alpha$.
To generate a W-state, we initialize the atom to be in the state $|g_0\rangle$ and all three photons to be in the state $|0\rangle$.
We then sequentially scatter three photons against the atom. Finally, we decouple the atom by projecting it to the state $|g_1\rangle$. The result is a W-type state 
$\eta\, |001\rangle + \xi \eta\, |010\rangle + \xi^2\, \eta\, |100\rangle$. To generate a GHZ state, we set the coupling constants $\gamma_0=\gamma_1$ such that $\xi=0,\,\eta = -1$.
  We first initialize the atom to be in the state $|g_0\rangle +|g_2\rangle$ and sequentially scatter three photons with input state $|0\rangle$. After each scattering, we rotate the atom state
 as $|g_0\rangle \rightarrow |g_1\rangle$ and $|g_1\rangle \rightarrow -|g_0\rangle$. At the final step, we decouple the atom by projecting it to the state  $|g_0\rangle +|g_2\rangle$. The result is a GHZ state $|000\rangle +|111\rangle$.

So far we show that one can generate a MPS by sequential scattering of single photons against an atom. 
Unlike previous works that use cavity QED systems for sequential generation of multi-photon MPS \cite{ssvcw, shwcs},  
where the different photon states are restricted to either the absence ($|{0}\rangle$) or presence ($|{1}\rangle$) of a photon in a given time bin, 
here the different photon states can be either different polarizations or frequencies. As result, the physical dimension of the MPS  can be higher than two. 
More importantly, there is a flexibility to rescatter an outgoing photon against the same atom. This flexibility enables one to generate PEPS that are more general than MPS.      
 
For illustration, we first present the procedure for generating a four-photon PEPS (Fig. \ref{fig1} (b)). 
Given four photons in the respective states $d^{[1]j_1}|j_1\rangle, \cdots, d^{[4]j_4}|j_4\rangle$, we sequentially scatter them against an atom with an initial state $I_{\alpha_0}|\alpha_0\rangle$ and 
perform unitary operations in the ground-state manifold of the atom
between two sequential scatterings. As discussed before, after the fourth photon scattering, we have a state 
${A}^{[4]i_4}_{\alpha_4 \alpha_{3}}{A}^{[3]i_3}_{\alpha_3 \alpha_{2}} {A}^{[2]i_2}_{\alpha_2 \alpha_{1}}{A}^{[1]i_1}_{\alpha_1 \alpha_{0}} I_{\alpha_0} |i_4i_3 i_2 i_1,\alpha_4\rangle$ with matrices $A$ defined in (\ref{Amat}). Now
instead of decoupling the atom,
we rescatter the first photon against the atom, followed by an unitary operation $\hat{U}^{[1]}$ on the atom. As a result, the atom-photon state
becomes $ U^{[1]}_{\beta_1\gamma_1}S_{\gamma_1\alpha_4}^{i_1j_1} {A}^{[4]i_4}_{\alpha_4 \alpha_{3}}
{A}^{[3]i_3}_{\alpha_3 \alpha_{2}} {A}^{[2]i_2}_{\alpha_2 \alpha_{1}}{A}^{[1]j_1}_{\alpha_1 \alpha_{0}} I_{\alpha_0} |i_4i_3 i_2 i_1,\beta_1\rangle$. We 
then rescatter the third photon against the atom and do an unitary operation on the atom. Finally, we decouple the atom by projecting it to the state $\mathbf{F}$ to end up with a four-photon entangled state: 
\begin{eqnarray}\label{tn1}
|\psi_4\rangle = F_{\beta_3}^*U^{[3]}_{\beta_3\gamma_3}S_{\gamma_3\beta_1}^{i_3j_3} 
U^{[1]}_{\beta_1\gamma_1}S_{\gamma_1\alpha_4}^{i_1j_1} {A}^{[4]i_4}_{\alpha_4 \alpha_{3}}{A}^{[3]j_3}_{\alpha_3 \alpha_{2}}
 {A}^{[2]i_2}_{\alpha_2 \alpha_{1}}{A}^{[1]j_1}_{\alpha_1 \alpha_{0}} I_{\alpha_0} |i_4i_3 i_2 i_1\rangle\,.
\end{eqnarray}
By defining tensors 
\begin{eqnarray}
T^{[1]i_1}_{\beta_1\alpha_4\alpha_1} \equiv U^{[1]}_{\beta_1\gamma_1}S_{\gamma_1\alpha_4}^{i_1j_1}{A}^{[1]j_1}_{\alpha_1 \alpha_{0}} I_{\alpha_0}\,,\,\,\,\,\,\,\,
T^{[3]i_3}_{\beta_1\alpha_3\alpha_2} = F_{\beta_3}^*U^{[3]}_{\beta_3\gamma_3}S_{\gamma_3\beta_1}^{i_3j_3} {A}^{[3]j_3}_{\alpha_3 \alpha_{2}}\,,
\end{eqnarray}
the state (\ref{tn1}) can be simplified to the form of 
\begin{eqnarray}
|\psi_4\rangle = {A}^{[4]i_4}_{\alpha_4 \alpha_{3}}T^{[3]i_3}_{\beta_1\alpha_3\alpha_2}
 {A}^{[2]i_2}_{\alpha_2 \alpha_{1}} T^{[1]i_1}_{\beta_1\alpha_4\alpha_1}|i_4i_3 i_2 i_1\rangle\,,
\end{eqnarray}
which is a PEPS with the graph representation of two triangles as shown in Fig.~\ref{fig1}(b).

The above procedure can be easily generalized to generate an arbitrary PEPS through sequential scatterings. Note that any PEPS state can be represented as an undirected graph. For any graph, 
the number of odd-degree nodes must be even.  
 If the graph contains zero or two odd-degree nodes, we can traverse the whole graph through a single Eulerian path.
In this case, we initialize the atom, do sequential scattering of photons following the visiting order of nodes in the Eulerian path, and finally  decouple the atom.
If the graph contains more than one pair of odd-degree nodes, we can traverse the whole graph through multiple disjoint Eulerian paths. For each Eulerian path, we initialize the atom at the begin of the path, do sequential scattering following the visiting order along the path and decouple the atom at the end of the path. 
In this process, we visit each edge only once. But we may visit a single node for multiple times, which corresponds to the operations 
where we re-scatter the same photon multiple times against the atom. Such a capability is unique to our scattering approach and is not available in previous approaches for using cavity QED systems for entangled state generation.

A related application is the generation of two-dimensional cluster states that are crucial in the quantum computing \cite{rb}.  
Recently, two-dimensional cluster states with square lattice structure have been generated using a cavity QED system \cite{pczl}. In our sequential scattering approach,
to generate the cluster state with arbitrary graph representation, 
we first design a waveguide QED system where two-photon sequential scattering 
process is identical to a controlled phase-flip gate \cite{dk, zgu}. We then initialize each photon to be in the state $|0\rangle+|1\rangle$ and perform sequential scatterings following the graph 
representation of the state.  In this way, we can generate cluster states with arbitrary graph representation. 

To summarize, we propose a sequential scattering approach to generate photonic MPS and PEPS using waveguide QED systems. 
We show that by performing sequential scattering of initially unentangled photons against an atom, we can generate arbitrary MPS with
the bond dimension equal to the number of the ground states of the atom in the waveguide QED system. Moreover,
by photon re-scatterings, we can generate PEPS with arbitrary graph representation.  
This approach provides significant more flexibility in generating MPS and PEPS as compared to previous approach, and points to the importance of waveguide QED systems for quantum state generations. 

This research is supported by AFOSR-MURI programs, Grant No. FA9550-12-1-0488 and  FA9550-17-1-0002.

%Similarly, the two-photon multi-channel S matrix is also a $2\times 2$ matrix $\mathbf{S}_{p_1p_2;k_1k_2}$ with each entry defined by
% $\left[\mathbf{S}_{p_1p_2;k_1k_2}\right]_{\mu\nu}\equiv\langle p_1,p_2, g_{\mu}|S|k_1,k_2,g_{\nu}\rangle$. $\mathbf{S}_{p_1p_2;k_1k_2}$ can be computed explicitly as
%\begin{eqnarray}
%\mathbf{S}_{p_1p_2;k_1k_2}&=&\frac{1}{2}\left(\mathbf{S}_{p_1k_1}\cdot \mathbf{S}_{p_2k_2}+\mathbf{S}_{p_2k_2}\cdot \mathbf{S}_{p_1k_1}+\mathbf{S}_{p_2k_1}\cdot \mathbf{S}_{p_1k_2}+\mathbf{S}_{p_1k_2}\cdot \mathbf{S}_{p_2k_1}\right)
%+\mathbf{C}\delta(p_1+p_2-k_1-k_2)\,,
%\end{eqnarray}

\end{document}